\documentclass{emulateapj}

\shorttitle{the Lupus Loop region \& \spear/\fims}
\shortauthors{Shinn et al.}

\newcommand{\cu}{ph cm$^{-2}$ s$^{-1}$ sr$^{-1}$ \AA$^{-1}$}
\newcommand{\lu}{ph cm$^{-2}$ s$^{-1}$ sr$^{-1}$}
\newcommand{\ncm}{cm$^{-3}$}

\newcommand{\Civ}{\ion{C}{4}}
\newcommand{\Niv}{\ion{N}{4}]}
\newcommand{\Hi}{\ion{H}{1}}
\newcommand{\Siii}{\ion{Si}{2}$^*$}
\newcommand{\ionSi}{Si$^+$}

\newcommand{\ionH}{H$^{+}$}
\newcommand{\dl}{$\lambda\lambda$}
\newcommand{\EBV}{$E(B-V)$}
\newcommand{\Rv}{$R_V$}
\newcommand{\fims}{\textit{FIMS}}

\newcommand{\spear}{\textit{SPEAR}}
\newcommand{\stsat}{\textit{STSAT--1}}

\begin{document}

\title{Diffuse Far-ultraviolet Observation of the Lupus Loop Region}

\author{Jong-Ho Shinn\altaffilmark{1}, Kyoung Wook Min\altaffilmark{1}, Chi-Na Lee\altaffilmark{1}, Jerry Edelstein\altaffilmark{2}, Eric J. Korpela\altaffilmark{2}, Barry Y. Welsh\altaffilmark{2}, Wonyong Han\altaffilmark{3}, Uk-Won Nam\altaffilmark{3}, Ho Jin\altaffilmark{3}, Dae-Hee Lee\altaffilmark{3}}

\email{jhshinn@kaist.edu}

\altaffiltext{1}{Dept. of Physics, Korea Advanced Institute of Science and Technology, 373-1 Guseong-dong, Yuseong-gu, Daejeon, 305-701, Republic of Korea}
\altaffiltext{2}{Space Sciences Laboratory, University of California, Berkeley, CA 94720, U.S.A.}
\altaffiltext{3}{Korea Astronomy and Space Science Institute, 61-1 Hwaam-dong Yuseong-gu Daejeon, 305-348, Republic of Korea}

\begin{abstract}
Diffuse far-ultraviolet (FUV) emissions from the Lupus Loop region (SNR 330.0+15.0) have been observed with \textit{Spectroscopy of Plasma Evolution from Astrophysical Radiation} (\spear), also known as \textit{Far-ultraviolet IMaging Spectrograph} (\fims).
We have detected several important ionic lines, including \Siii{}, \Civ{}, and N IV], which characterize the warm, hot ionized gas in this region.
The spatial variations in the line intensities of \Siii{} and \Civ{} have also been studied in comparison with X-ray and dust observations.
The result shows that they originate from the interface between the hot gas seen in the X-ray and the cooler \Hi{} shell with which dust is associated.
The interface is rather diffuse, and the gases with different temperatures seem to co-exist in this region.
A shock may exist upfront in the interface, but its velocity should be very small as no shock-related distinguished feature is seen in H$_{\alpha}$.
\end{abstract}

\keywords{ISM: individual (Lupus Loop) --- supernova remnants --- ISM: clouds --- ISM: bubbles --- ISM: evolution}

\section{Introduction} \label{intro}

The Lupus Loop (SNR 330.0+15.0) is a large nonthermal radio source ($\diameter\sim270\arcmin$) that lies near the supernova remnant (SNR) SN 1006 (cf.~Fig.~\ref{subrg}).
In the radio-continuum domain, it shows a diffuse appearance, low surface-brightness \citep{mil71}, and a globally uniform-and-intense polarization \citep{mil74}; however, there is no large-scale nebulosity in the optical domain \citep{gre04}.
Despite its small angular separation from SN 1006 ($\sim 2 \fdg 4$), it is thought not to be associated with the supernova (SN) because it requires an unrealistically high mean expansion-velocity ($\sim0.1-0.5c$) \citep{mil71,ste77}.
The distance to the Lupus Loop ($d_L$) is still uncertain: it varies from 0.17 kpc \citep{ola81} to 0.5 kpc \citep{cla76} and even to 1.2 kpc \citep{lea91}.

The Lupus Loop region was studied in X-ray and radio, and the results showed a complex morphology surrounding the region.
For example, \citet{rie80} observed two enhanced $0.15-3.0$ keV X-ray regions around the Lupus Loop.
\citet{col82} observed two concentric \Hi{} shells and found that they are well-anticorrelated with the enhanced X-ray regions. (cf. Fig.~\ref{subrg}).

The density of the ambient interstellar medium around the Lupus Loop ($n_{a}$), the age ($Y_L$) and the temperature ($T_L$) of the Loop, and the initial total energy of the SN explosion ($E_0$) were derived from several X-ray observations \citep{win79,gro79,too80,lea91}: $n_a$, ranging from 0.02 to 0.1 \ncm{}, was somewhat low considering $n_{H}=0.07-0.5$ \ncm{}, the \Hi{} density at a similar galactic altitude \citep{dic90}.
\citet{col82} also estimated $n_a$ ($\sim0.13$ \ncm) from the column density of the inner \Hi{} shell.
Estimates of $Y_L$, ranging from $20,000$ to $50,000$ yr based on the X-ray observations, are consistent with \citeauthor{mil71}'s \citeyearpar{mil71} suggestion, based upon its diffuse appearance and low surface-brightness in radio, that the Lupus Loop is an old SNR.
Estimates for $T_L$ and $E_0$ are on the order of $10^6$ K and $10^{50}-10^{51}$ erg, respectively.

We present in this paper the results of FUV emission line studies using the dataset from a new spectro-imaging observation of the Lupus Loop region.
We focus on the spatial variation in the intensities of \Siii{} $\lambda1533$ and \Civ{} \dl1548,1551 emission lines and interpret the results in comparison with the features seen in infrared (IR) and X-ray images.

\section{Observation and Data Reduction}

The FUV observation was performed by \spear, also known as \fims, aboard the first Korean Science and Technology SATellite, \stsat.
The {\it SPEAR/FIMS} mission employs an imaging spectrograph which is optimized for the observation of the diffuse emission lines in the FUV domain.
The L channel of the spectrograph covers 1350--1700 \AA{} with a resolving power of $\lambda/\Delta\lambda \sim$ 550.
The field of view is $7.4^\circ \times 4.3'$ with a 5--10$'$ imaging resolution along the slit.
The spectral half-energy line width, averaged over the angular field, is 3.2 \AA. The instrument, its in-orbit performance, and the data analysis procedures are described in \citet{ede06i}.

The Lupus Loop region was observed during a sky survey between 6 May 2004 and 27 May 2004.
Some of the observations were made with the shutter reduced to 10\% of the full slit length.
Overall the exposure time is 1 to 6 s.
The detector dark background of $0.03\pm0.01$ counts s$^{-1}$ \AA$^{-1}$ was subtracted from the raw count rates.
We followed the general analysis procedures described in \citet{ede06i}.
As our concern is mainly the diffuse emission in and around the Lupus Loop, it is important to remove nearby bright point-sources.
We have, therefore, excluded the pixels of which the wavelength-averaged surface-brightness is greater than 8,000 CU (CU$\equiv$ \cu; continuum unit).
The criterion was set in view of the results of \citet{sch01}, in which the maximum intensity of a dust-scattered FUV-continuum was found to be about 5,000 CU near dust clouds and the Galactic plane.
Spectra were created with 1.5\AA{} bin and then smoothed with a 4.5 \AA{}-wide box-car to improve the signal-to-noise ratio of each bin.

For comparisons of our results in FUV observation with the features seen in X-rays and IR domain, which characterize the $10^6$ K hot gas and cold dust, respectively, we have employed the archival data of the ROSAT All Sky Survey (RASS) maps \citep{sno95} and the Schlegel-Finkbeiner-Davis (SFD) 100 $\mu$m IR map \citep{sch98}.
These maps were extracted using the SkyView Observatory \citep{mcg98}.

\section{Analysis and Results} \label{anr}
\begin{figure}
\includegraphics[angle=90,scale=0.6]{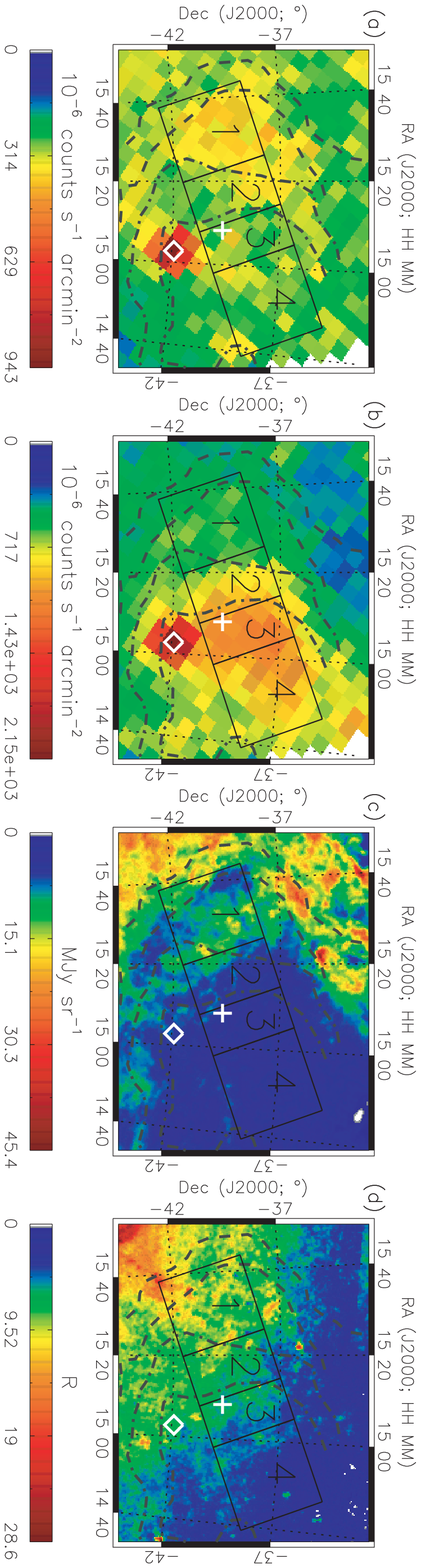}
\caption{\textsf{The Lupus Loop region in different wavelengths.} We select four subregions based on these images in different wavelength domains.
The two dash-dotted curves and the dashed curves represent the boundaries of the inner and outer \ion{H}{1}{} shells identified by \citet{col82}, respectively.
The maps are centered at ${\alpha}=$ (15$^{h}$ 20$^m$ 0.0$^s$, ${\delta}= $ --37\arcdeg{} 30\arcmin{} 0.0\arcsec) in J2000.
`$\Diamond$' and `+' indicate SN 1006 and the projected center of the Lupus Loop, respectively.
(a) Rosat All Sky Survey (RASS) 1.5 keV map \citep{sno95}, (b) RASS 0.25 keV map \citep{sno95}, (c) Schlegel-Finkbeiner-Davis (SFD) 100 $\mu$m map \citep{sch98}, and (d) H$_{\alpha}$ map \citep{fin03} \label{subrg}}
\end{figure}

We have divided the observed Lupus Loop region into four subregions based on the X-ray, IR, and H$_{\alpha}$ maps, as shown in Fig.~\ref{subrg}.
The core of Region-1 is occupied by hot gas, as seen in Fig.~\ref{subrg}a of the RASS 1.5 keV map.
The two \Hi{} shells identified by \citet{col82} surround Region-1, as indicated in Fig.~\ref{subrg}a.
Region 2 coincides, more or less, with the inner \Hi{} shell while Region-3, located inside the inner shell, shows the most intense X-ray in RASS 0.25 keV map.
Region-4 is a test region for comparison with others.
The SFD 100 $\mu$m map (Fig.~\ref{subrg}c) shows that Region-1 also includes dust near the northwestern boundary of the region.
The same dust layer generally coincides with the inner \Hi{} shell in other parts. Hence, dust is also seen in Region-2.
In Fig.~\ref{subrg}d, it can be seen that H$_{\alpha}$ is prominent only around the hot gas in Region-1 and it diffuses away toward Region-2 and -3.
As these maps indicate, hot gas coexists with cooler gas and dust in the broad region of Lupus Loop. 
Hence, we expect cooling processes to occur in their interaction regions. 
The purpose of the present work is to study such processes using the new FUV data obtained by \spear/\fims.

\begin{figure}
\includegraphics[scale=0.9]{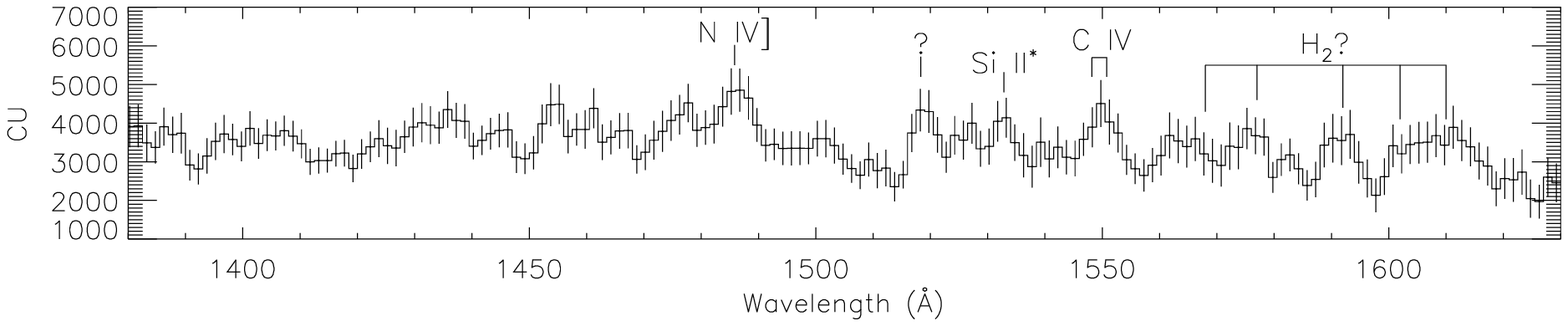}
\caption{\textsf{Identification of emission lines in the total spectrum.}  The total spectrum of all subregions, binned in 1.5 \AA, is smoothed with a box-car whose width is $3\times$(bin size). \Siii{} $\lambda$1533, \Civ{} \dl1548,1551, and \Niv{} $\lambda$1486 are identified. \label{tspec}}
\end{figure}

Fig.~\ref{tspec} shows the total spectrum of all subregions in the 1380--1630 \AA{} band.
Some of the important atomic emission lines, such as N IV $\lambda$1486, \Siii{} $\lambda$1533, and \Civ{} \dl1548,1551, are easily identified.
The peak around $\lambda$1409 might represent the blending of the close O IV and S IV lines.
\ion{Si}{2} $\lambda$1527, a companion of the doublet \Siii{} $\lambda$1533, is not seen in the spectrum.
We believe it is due to the resonant scattering since \ion{Si}{2} $\lambda$1527 is the transition to the true ground state and is likely to be optically thick in a neutral and warm ionized medium.
It should be noted that the line was not detected in a similar study for the Orion-Eridanus superbubble \citep{kre06}.
The spectrum also hints of the existence of molecular hydrogens.

In order to study the spatial variation of FUV emission charateristics, we have extracted the spectrum from each subregion around \Siii{} $\lambda$1533 and \Civ{} \dl1548,1551, which represent $10^{4}$ K and $10^5$ K, respectively, and have estimated the intensities of these lines using a $\chi^2$ minimization method \citep{kri94}.
The \Siii{} emission line was fitted with a single Gaussian profile while the \Civ{} emission lines were fitted with two Gaussian profiles with their line ratio $\lambda1548:\lambda1551=2:1$.
The center wavelengths of the profiles were all fixed, and we assumed a linear continuum with its level as a free parameter.
An extinction correction was made using the standard interstellar extinction curve \citep[with \Rv= 3.1]{car89}.
We have used \EBV=0.03, as derived by \citet{cra00} for the outflowing material from the Sco-Cen OB association \citep{cra91,deG92}, assuming this outflowing material resides in the line of sight toward the diffuse emission region of our concern.
We believe this assumption is reasonable because the Lupus Loop region is located amid the angular coverage of the Sco-Cen OB association and the distance to the outflow region is only \emph{$\sim$100 pc} \citep{cra00} while that to the Lupus Loop is \emph{$\ga$170 pc}.

\begin{figure}
\includegraphics[angle=90,scale=0.5]{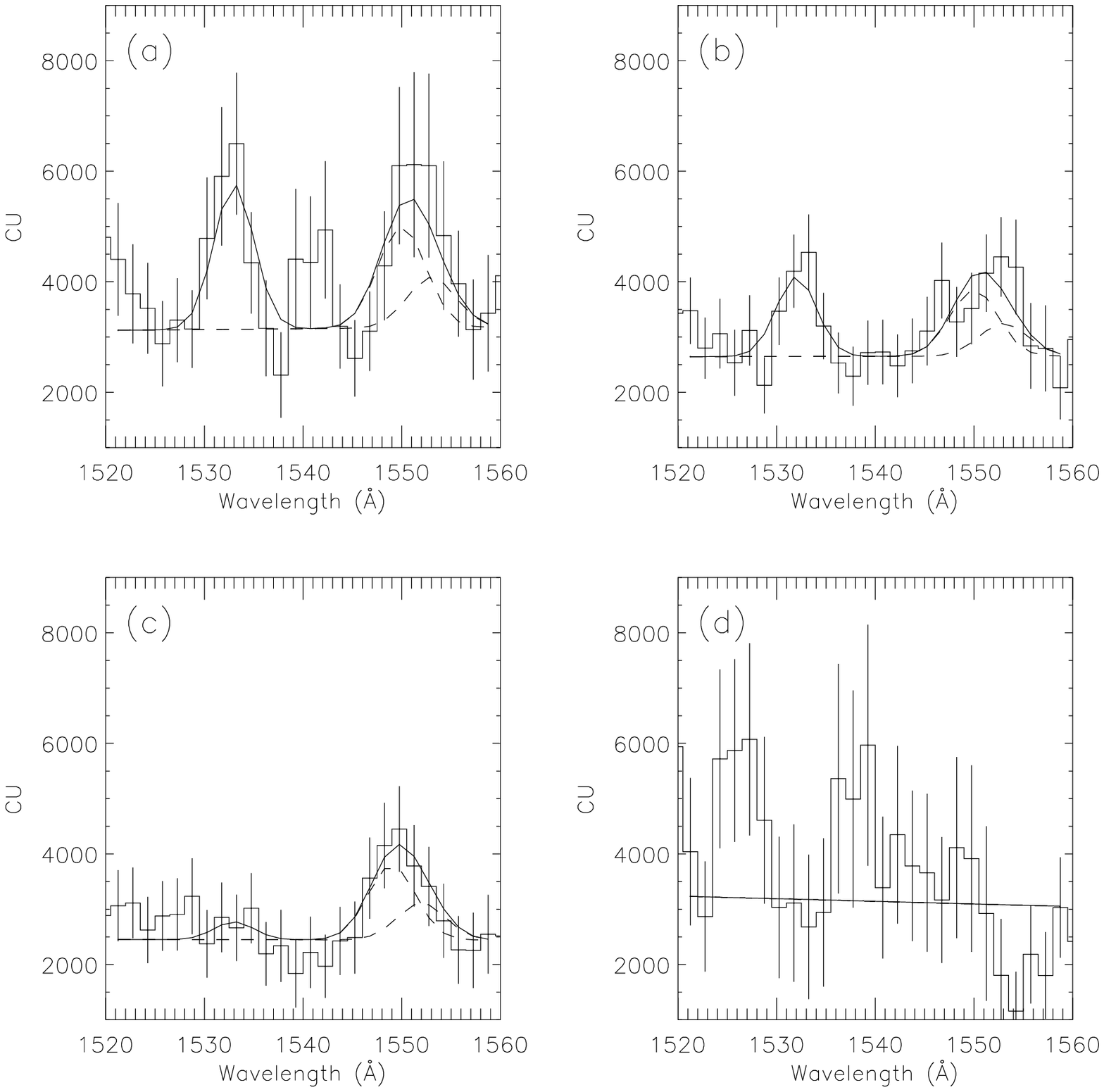}
\includegraphics[angle=90,scale=0.5]{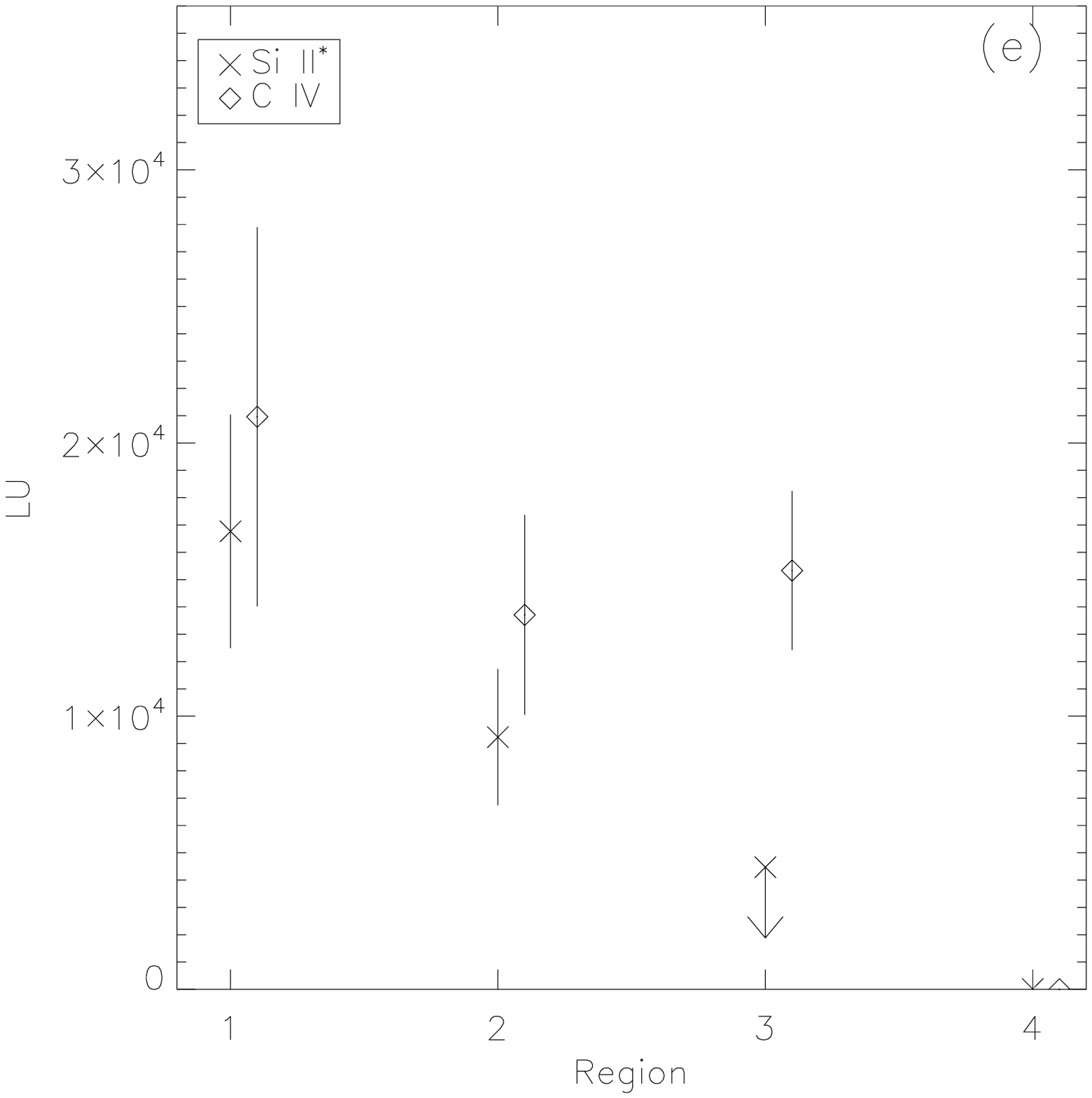}
\caption{\textsf{Fitting of \Siii{} $\lambda$1533 and \Civ{} \dl1548,1551 for each subregion.} The histogram, solid-line, and dashed-line in (a)--(d) represent the L-channel spectrum with 1-$\sigma$ error bar, fitting lines, and the two components of \Civ{} \dl1548,1551 fitting lines, respectively.
In (e), where the line intensities of \Siii{} and \Civ{} are shown for each subregion, the \Civ{} mark is slightly shifted to the right to avoid the overlapping of the 1-$\sigma$ error bar.  \Siii{} in Region-3 and -4 are given with 90\% upper limits, as indicated by arrows.
No acceptable fitting was found for both \Siii{} and \Civ{} in Region-4 with finite intensity. \label{result}}
\end{figure}

Fig.~\ref{result}a--d show the spectra and fitting results for each subregion.
It can be seen that \Siii{} is prominent in Region-1 and -2 while its intensity decreases as we move from Region-2 to Region-3.
There are no noticeable \Siii{} or \Civ{} lines in Region-4.
We plot in Fig.~\ref{result}e the estimated \Siii{} and \Civ{} line intensities for each subregion in LU ($\equiv$\lu; line unit).
Again, \Siii{} is detected in Region-1 and -2 with 3-$\sigma$ confidence while \Civ{} is detected with the same significance in Region-1, -2, and -3.
The relative intensity ratio of \Civ{} to \Siii{} increases from Region-2 to Region-3, and neither of the lines is detected in Region-4 with 3-$\sigma$ confidence.

\section{Discussion}

Based on the \Hi{} observations, together with the X-ray results, \citet{col82} conjectured that the Lupus Loop was formed by a supernova explosion inside a bubble of pre-existing stellar wind.
Each of the two \Hi{} shells represents the boundary of the stellar wind and the supernova-shocked gas, respectively, while the stellar wind between the two \Hi{} shells and the innermost supernova-shocked gas are manifested as the regions of enhanced X-rays.
The Lupus Loop region has been observed since then in other wavelength domains as well as in X-rays with higher spectral and spatial resolutions.
Fig.~\ref{subrg} summarizes the recent observations of the region.
The two RASS maps of Fig.~\ref{subrg} a and b, which represent 1.5 keV and 0.25 keV, respectively, indicate that the two hot gas regions may have different temperatures as Region-1 is brighter than Region-3 in the 1.5 keV map, while Region-3 is brighter than Region-1 in the 0.25 keV map.
Following \citet{sno97}, in which the intensity ratios of the RASS bands are translated into the corresponding hot gas temperatures, we estimate the temperatures to be $T_{R1}\sim4.9\times10^6$ K and $T_{R3}\sim4.5\times10^6$ K while the associated uncertainties are large enough for the temperatures of these two regions overlap with each other.
Fig.~\ref{subrg}c, the SFD 100 $\mu$m map, shows clumps of dust outside the outer \Hi{} shell and a rather broad layer along and just outside the inner \Hi{} shell.
The upper left region of the figure is the Lupus Cloud, and the lower left corner merges into the galactic plane.
On the other hand, H$_{\alpha}$ intensity, shown in  Fig.~\ref{subrg}d, generally decreases more or less monotonically from the galactic plane toward Region-4, except around the hot gas in Region-1.
Region-4 is devoid of H$_{\alpha}$.
The region near the upper boundary of the figure also shows low H$_{\alpha}$ intensity where the IR map shows the existence of dust.

The present FUV study provides additional information to the features described above.
In Region-1, where both hot gas and dust exist, both \Siii{} and \Civ{} are detected with 3-$\sigma$ confidence.
We believe the corresponding ionized Si originates from dust in this region through thermal evaporation by the hot gas or photo-ionization by the interstellar ultraviolet radiation field.
Region-2, representing the inner \Hi{} shell, is the interface between the dust layer in Region-1 and the hot gas in Region-3, and both \Siii{} and \Civ{} are detected in this region with 3-$\sigma$ confidence.
This is reasonable as they represent $\sim10^4$ K and $\sim10^5$ K gases, respectively.
In Region-3, filled with hot gas and located farther away from the interface, it is natural that \Civ{}, emanating from the cooling zone around the hot gas, is more prominent than \Siii{}.
In Region-4, neither \Siii{} nor \Civ{} is detected as this region is expected to be devoid of dust and mainly consists of tenuous hot gas.

It is worthwhile to note that a study similar to the present one has been conducted for the Orion-Eridanus Superbubble \citep{kre06} using \spear{} data.
Both \Siii{} and \Civ{} were found in their study from the thermal interface, at which H$_{\alpha}$ also peaked.
On the other hand, H$_{\alpha}$ does not show a notable feature in the interface region, Region-2 and -3, in the present case.
\citet{col82} estimated the expansion velocity of the inner \Hi{} shell to be $23\pm3$ km s$^{-1}$, which corresponds to the post-shock temperature of $T_{ps}=3{\mu}m_HV_{s}^2/16k\simeq1.2\times10^4$ K, where $\mu$ is the molecular mass in units of the hydrogen atom mass ($m_H$).
Considering the ionization energies of \ionSi{} and \ionH, 8 eV and 14 eV, respectively, this low temperature might explain the absence of prominent H$_{\alpha}$ while \Siii{} can still be detected.

\section{Summary}
Using the new FUV dataset obtained from {\it SPEAR/FIMS}, we have studied the spectral characteristics of the Lupus Loop region.
Our result shows that \Siii{} and \Civ{} emission lines emanate from the interface between the hot gas and the cooler \Hi{} shell with which a dust layer is also associated. \Siii{} is more prominent closer to the region while \Civ{} farther away from it.
This result is consistent with other observations in X-rays, IR, and radio, each of which shows distinctive features.
The interface is rather broad and a shock, whose structure is not clear in the FUV result, should have a very low velocity if it exists.

\acknowledgments
\emph{SPEAR/FIMS} is a joint project of the Korea Advanced Institute of Science and Technology, the Korea Astronomy and Space Science Institute, and the University of California at Berkeley, funded by the Korea Ministry of Science and Technology and the U.S. National Aeronautics and Space Administration Grant NAG5-5355.
The authors appreciate Dubner for providing the \Hi{} image. 
They are also grateful for valuable suggestions of the referee.
Some of the results in this paper have been derived using the HEALPix \citep{gor05etal} package.

\bibliographystyle{apj}

\clearpage

\clearpage

\clearpage

\end{document}